\documentclass[%
 ,twocolumn%
 ,secnumarabic%
,amssymb, amsmath,nobibnotes, aps, prl]{revtex4}
\usepackage{docs}%
\usepackage{bm}%
\usepackage{graphics}%
\expandafter\ifx\csname package@font\endcsname\relax\else
 \expandafter\expandafter
 \expandafter\usepackage
 \expandafter\expandafter
 \expandafter{\csname package@font\endcsname}%
\fi

\begin{document}

\title{Trapping radioactive $^{82}$Rb in an optical dipole trap and evidence of spontaneous spin polarization}

\author{D. Feldbaum, H. Wang, J. Weinstein, D. Vieira, and X. Zhao}
\email{xxz@lanl.gov}
\affiliation{Los Alamos National Laboratory, Los Alamos, NM 87545}
\date{\today}%

\begin{abstract}
Optical trapping of selected species of radioactive atoms has great potential in precision measurements for testing fundamental physics such as EDM, PNC and parity violating $\beta$-decay asymmetry correlation coefficients. We report trapping of $10^4$ radioactive $^{82}$Rb atoms $(t_{1/2}=75$ s) with a trap lifetime of $\sim$55 seconds in an optical dipole trap. Transfer efficiency from the magneto-optical trap was $\sim$14\%.  We further report the evidence of spontaneous spin polarization of the atoms in optical dipole trap loading. This advancement is an important step towards a new generation of precision  $\mathbf{J}-\mathbf{\beta}$ correlations measurements with polarized $^{82}$Rb atoms.
\end{abstract}

\maketitle

Recent advancements in atom trapping have resulted in exciting breakthroughs in atomic, molecular and optical physics, with the realization of diverse phenomena such as Bose-Einstein condensation \cite{BEC1, BEC2, BEC3, BECbook}, Fermi-Degenerate gases \cite{Jin} and fermion superfluidity \cite {Ketterle}, as well as ultracold plasmas \cite {Michigan, Killian} and atomic clocks.

The technique of trapping ultracold atoms also has great potential for precision measurements because it can provide an ideal sample in a well controlled environment. There are several attempts and ongoing efforts in $\beta$-recoil measurements on $^{38m}$K $(t_{1/2}=0.9$ s)  \cite{TRIUMF} and $^{21}$Na $(t_{1/2}=21$ s) \cite {lbl} in a magneto-optical trap (MOT),  $\beta$-spin correlation of $^{82}$Rb in a time-orbiting potential(TOP) trap \cite{LANLPNCinTOP}, Fr PNC \cite{StonyBrooks, PisaLegrano} and Ra EDM \cite{Lu, Guest}.  Yet no group has demonstrated the trapping of short lived radioisotopes in an optical dipole trap, largely due to small number of atoms in the primary MOT and poor transfer efficiency from the MOT to the dipole trap.

\begin{figure}
\includegraphics{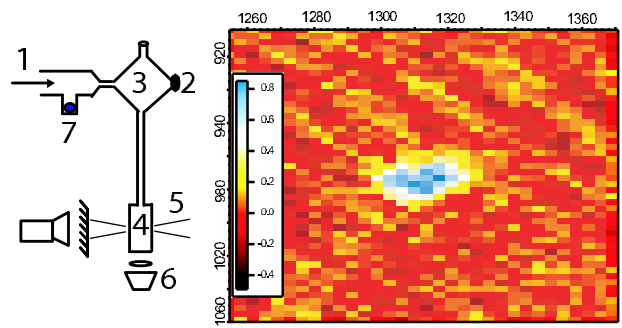}
\caption{\label{setup}The experimental setup and an image of $4\times10^3$ $^{82}$Rb atoms in a dipole trap. The $^{82}$Rb ions are incoming from the mass separator (1), and focused on the Zr foil (2). The atoms are then released from the foil as neutrals, captured in the primary MOT (3), and are optically pushed into the experimental chamber (4). There the atoms are recaptured in a MOT, and are, finally, transferred into an optical dipole trap (5). All the data in this paper has been obtained using a photodiode (6) and with an on-axis CCD camera. Stable atoms are obtained from an oven (7).
The absorption image is taken with the TOF of $50~\mu$s. The color scheme denotes the optical density as shown on the scale. One pixel corresponds to 3.7 $\mu$m. The yellow fringes originate from the parallel walls of the trapping cell. The FORT beam intensity is $2.3\times 10^5$ W/cm$^2$, and the calculated trap depth is $340~\mu$K.}
\end{figure}

Radioactive atoms confined in a far-off-resonance dipole trap (FORT) have many intrinsic advantages for fundamental symmetry experiments, providing a highly polarized, point-like sample with minimal perturbation from the environment which can be well characterized. It is an ideal system for studying parity violating $\beta$-decay of spin-polarized protons. Parity violation was first suggested by Lee and Yang \cite{LeeYang}, and subsequently discovered in 1957 by Wu \emph{et al.} \cite{Wu}, in the beta decay of polarized $^{60}$Co.  Today, parity violation is encompassed by the standard model (V-A) interaction between leptons and quarks.  Nonetheless, the nature of these helicity couplings is derived from empirical measurements and the standard model offers no fundamental understanding of the origin of these symmetries and how they become broken at the energy scales probed by modern experiments.  Low energy physics experiments that exploit nuclear beta decay continue to offer a means to probe the fundamental origin of parity violation and, more generally, the helicity structure of the weak interaction \cite{Deutsch}.  With advantages already mentioned above, the use of pure optical trap combined with optical pumping will enable a state-of-the-art $\beta$-asymmetry measurement. The challenges to overcome is to trap sufficient number of atoms with long enough trapping lifetime and precise measurement on the atomic/nuclear polarization. In this paper, we report the first trapping of radioactive $^{82}$Rb atoms with good trapping efficiency and long lifetime and optical manipulation of the spin states in a dipole trap. The requirement to obtain the lifetime of trapped atoms comparable with the nuclear decay lifetimes necessitates the use of a separate science chamber, and a large percentage of the atoms is lost in the process of MOT-to-MOT transfer. Later we plan to transfer the $^{82}$Rb atoms directly from the primary MOT into an optical tweezer and transport the atoms to a science chamber for a $\beta$-asymmetry measurement. We expect an order of magnitude improvement in atom number.

 The radioactive atom trapping system consists of an existing double-MOT coupled to a mass separator (Fig. \ref{setup}). The trapping of $^{82}$Rb in the primary MOT has been described in detail previously \cite {LANL82Rb, LANLPNCinTOP}. Briefly, 10 to 50 mCi of $^{82}$Sr ($t_{1/2}=26$ d) is loaded in the ion source of the mass separator to be used as a parent source for $^{82}$Rb.
 $^{82}$Rb is surfaced ionized, accelerated to 20 keV, mass separated, and is collected for about 3 half-lifes ($\sim200$ s) on a 5 mm diameter Zr foil located inside the primary trapping cell. When induction heated to $\sim$$ 800^o$C, the Zr releases the accumulated $^{82}$Rb as neutral atoms with $\sim$50\% release efficiency as measured by $\gamma$ counts from the foil. The atoms released by the foil are then laser cooled and trapped in a magneto-optic trap (MOT). The trapping cell is coated on the inside with a non-stick trimethomethoxysilane (SC-77) dry film which enhances the trapping efficiency by two or more orders of magnitude. The coating can last up to a year in good vacuum ($\sim10^{-9}$ Torr), but degrades when exposed to the Rb vapor or high temperatures, such as resulting from the foil heating. Our coating degraded by a factor of $\sim$20 over ~6 months of experiment. 

The trapping light is detuned $-15$ MHz from the $F=3/2 \rightarrow F'= 5/2$ trapping transition of the $^{82}$Rb by locking the Ti-Sapphire laser to the $F=3 \rightarrow F'=3,4$ crossover of the $^{85}$Rb and frequency shifting $-536$ MHz using a double-pass AOM. The repump beam is produced as the sidebands on the trapping beam by an EOM at 1.472 GHz. To detect the MOT fluorescence, we modulate that EOM at 4 kHz and detect the modulated fluorescence with an avalanche photodiode and a lock-in amplifier.  In our current system, the trapped $^{82}$Rb atoms are transferred into a science chamber through a 100 cm long tube with magnetic confinement along the tube axis. The science chamber is a high quality rectangular quartz cell. A combination of ion and Ti sublimation pumps are used to achieve a vacuum of better than $10^{-11}$ Torr in the second chamber. This gives a MOT2 lifetime of $\sim200$ s.   We are able to obtain as many as $10^5$ $^{82}$Rb atoms in the MOT2 after a single MOT1-MOT2 with a transfer efficiency of no more than $\sim$$10\%$ due to long transfer distance.  In the future, this MOT1 to MOT2 transfer step will be eliminated by using an optical tweezer to get more atoms in the dipole trap to obtain statics for $\beta$ decay events.

The transferred atoms are re-trapped in a second MOT and then are prepared for loading into a dipole trap.  We use a high power single frequency solid state laser with $\lambda=1030$ nm for the dipole trap. The laser beam is sent through an AOM operating at 40 MHz.  Up to 80\% of the light intensity is deflected into the first order, which is coupled into a single-mode fiber. The beam from the fiber is then focused to a $\sim30~\mu$m $1/e$ diameter waist after passing through a series of lenses, polarizing cube and waveplates and is overlaped with the MOT cloud using a single dichroic mirror.
For the optical lattice, the beam passes through a second achromat with the same focal length of 150 mm  and reflected back by a YAG mirror at a normal incidence. The YAG mirrors used are transparent at $780$ nm, allowing us to perform the absorption imaging of the MOT using a probing beam along the YAG laser beam on the camera. The fluorescence of the MOT is also collected onto a second CCD camera located at 90 degree from the absorption camera.

The FORT loading process consists of compressed and detuned dark MOT stages to achieve a higher density of cold atoms. The increase of the MOT magnetic field gradient by a factor of 4 and detuning the laser frequency by an additional $-18$ MHz results in the improved dipole loading efficiency of up to $\sim$14\%.  After a "FORT holding time" delay we switch off the FORT light and absorption imaging the atoms on the CCD using a $50~\mu$s probe beam to determine the absolute number and temperature of the trapped atoms. To image atoms in the lower hyperfine state, a 100 $\mu$s repump pulse was used before the probe. For relative measurements, we use the fluorescence of the atoms re-captured in a MOT without imaging.

We optimize the double MOT system and the FORT loading using $^{85}$Rb from a Rb getter. Once optimized, $^{82}$Rb can be trapped without any change in the trap alignment and timing. A necessary step in obtaining the FORT is the precise overlap of the FORT beam with the location of the compressed MOT with the help of the camera focused on the latter.  Our probe absorption beam is collinear with the axis of the trap, therefore in the case of the optical lattice we were imaging the total number of atoms in all the occupied sites at once. We observe the separation of the lattice into two individual dipole traps if we misaligned the retro-reflecting mirror of the lattice. 

We use the CCD camera image to establish the $1/e$ diameter of the trapping beam at 30 $\mu$m. We have further verified our calculations by measuring the parametric resonance frequencies for the atoms trapped with a 1.6 W beam.

For a successful $\beta$-asymmetry measurement, we have to achieve good transfer efficiency from the MOT to the dipole trap with a lifetime comparable to the nuclear decay lifetime. Further, the trapped atoms have to be polarized to the stretched state and the degree of atomic/nuclear polarization has to be measured to a high precision.

\begin{figure}
 \includegraphics{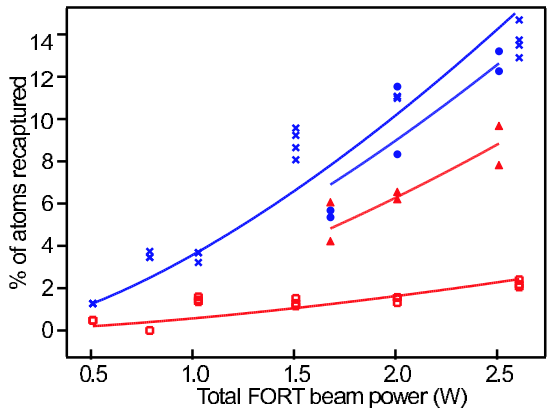}
 \caption{\label{TransEffic} The efficiency of MOT-to-FORT transfer of Rb atoms as a function of the FORT beam light intensity, as measured by MOT re-trap. The FORT beam remains on through out the experiment. Red squares: $^{85}$Rb transferred from "bright" MOT; red triangles: $^{82}$Rb transferred from "bright" MOT; blue circles: $^{82}$Rb transferred from dark MOT; blue crosses: $^{85}$Rb transferred from the dark MOT. The fits are derived from the $P^{3/2}$ relation \cite{LANLparametric}. Total FORT power of 1 W approximately corresponds to a $200~\mu$K trap depth.}
\end{figure}

As demonstrated in earlier dipole work \cite{LANLparametric}, we expect improved trapping
efficiency with increased laser power. Fig.~\ref{TransEffic}  shows the transfer efficiencies 
as a function of dipole laser power for both the dark compressed MOT and bright compressed MOT (repump left on during compression).  Much improved efficiency was observed for the dark compression which peaks at $\sim$14\% with approximately 2 W power (trapping depth $=425~\mu$K). The dependence of the MOT-FORT transfer efficiency on the FORT power $P$ is consistent with $P^{3/2}$ relation as discussed in \cite{LANLparametric}.

\begin{figure}
 \includegraphics{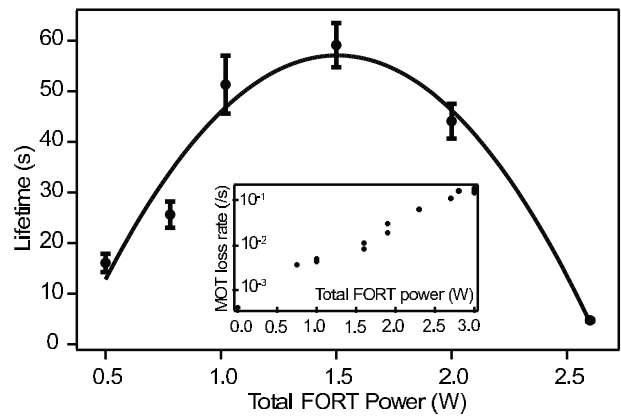}
 \caption{\label{LifetVSpower} The dependence of the lifetime of $^{85}$Rb atoms in the FORT on the intensity of the FORT light. The insert shows the detrimental effect of the FORT light on the lifetime of the MOT. Total FORT power of 1 W approximately corresponds to a $200~\mu$K trap depth.}
\end{figure}

The lifetime of the dipole trap as the function of laser intensity and power is shown in Fig. \ref{LifetVSpower}.   It peaks as power goes up to 1.5 W and start to drop at high power. The drop in trap lifetime at high power is due to the increased fiber output power instability.  This increased noise was observed on the power spectrum below 5 kHz when the fiber output is more than 2 W. We also found that FORT laser light has a pronounced effect on the MOT lifetime. The insert in Fig.~\ref{LifetVSpower} shows the loss rate of the atoms from the MOT as a function of the FORT light power.  The loss is a one body process and the exponential power dependence is completely different from the dipole trap loss. We think it is due to the level-shifts caused by the dipole trapping laser on the trapped atoms. The combination of the red shift of the ground $5S_{1/2}$ state, and of the blue shift of the $5P_{3/2}$ state results in an effective trapping laser detuning  of -7.8 MHz per 100 $\mu$K \cite{polarizabilities}, or, approximately, -16.6 MHz per W of our total FORT laser power in our experimental conditions. This detuning leads to an increased atom loss from the MOT. This effect warrants further investigation.

\begin{figure}
\includegraphics{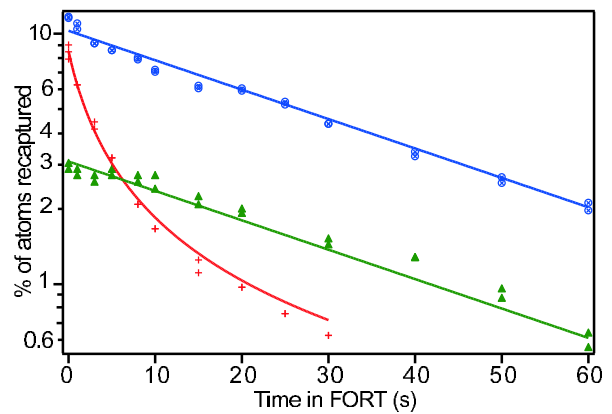}
\caption{\label{StateLifetimes} The lifetimes of the $^{85}$Rb atoms transferred from MOT into the FORT. Both atoms loaded from the dark MOT (blue circles) and from the bright MOT (green triangles) without an intermediate repumping step exhibit the same one-body decay lifetime of $\sim 37$ s (blue and green line fits), however the decay is two-body if the atoms are illuminated by a 100 $\mu$s repumping light after they are loaded into the FORT from a dark MOT (red crosses and red line fit).}
\end{figure}

The dipole trap loss also depends on the spin composition of the trapped atoms. Fig. \ref{StateLifetimes}  shows the decay of $^{85}$Rb atoms in the trap. When the atoms are loaded directly from the MOT in either the upper (bright MOT) or lower (dark MOT) hyperfine state, the decay is single exponential, a signature of one body decay. 
This indicates that the atoms are polarized to a single spin state, $F=3, m_F=3$ or $-3$ for the upper hyperfine state and  $F=2, m_F=2$ or $-2$ for the lower, resulting in neglible  two-body decay rate. The fact that the atoms are polarized without conventional optical pumping is a pleasant surprise and requires further investigation. We think it might be due to the spontaneous spin polarization during the dipole trap loading. The spontaneous spin polarization has been studied in Cs vapor \cite{Lamoreaux}, \cite{UW}, and investigated theoretically for the case of BECs \cite{Chinese}. Fig. \ref{StateLifetimes} also shows that when atoms are depolarized with a repump beam, the decay becomes a clearly two-body process. If the atoms are loaded into an optical lattice instead of a single-beam trap, the trap decay also has the two-body signature.

In summary, we have demonstrated the first ever trapping of short lived radioactive isotope in optical dipole trap with a transfer efficiency of $\sim$14\% and a long trapping lifetime of 55 s. This is an important step towards the beta asymmetry measurement in an optical dipole trap and suggests that sufficient number of atoms and decay statics can be obtained if we load atoms in the primary MOT to the dipole trap and transfer the atoms to a science chamber.  Once atoms are moved to the science chamber in a dipole trap, they can be polarized to the stretched state, and their polarization measured, e.g., with Faraday rotation signal. To measure the degree of polarization to 0.1\% accuracy, optical imaging methods could also be employed to sample the fraction of atoms (1\%) that are trapped but not in the fully-aligned state with 10\% precision. This can be accomplished with $10^5$ $^{82}$Rb atoms in the dipole trap.
The off resonance scattering at 1030 nm is on the order of 1 Hz depending on the trapping parameters.  This may require the 1030 nm dipole laser to be circularly polarized relative to the uniform bias field, so once the atoms are polarized to the stretched state, the off-resonance scattering maintains the polarization.  Angular distribution  can be measured with multiple detectors.  Naturally, one can also reduce this scattering rate to the order of 0.001 Hz with a CO$_{2}$ laser dipole trap, this would enable the application of a rotating bias field without any appreciable effect on the sample polarization, so that a complete mapping of angular distribution can be measured with a single detector \cite{LANLPNCinTOP}.   

We thank Wayne Taylor, Jason Kitten, Cleo Naranjo, and the Chemistry Division hot cell for providing $^{82}$Sr isotope, and Andrew Hime, Steve Lamoreaux, and Eddy Timmermans for helpful discussions related to $\beta$-asymmetry and spin polarization. This work was supported in large part by the Laboratory Directed Research and Development program at Los Alamos National Laboratory, operated by the Los Alamos National Security, LLC for the NNSA U. S. Department of Energy.
\\
\\

\end{document}